\documentclass[aip,apl,reprint]{revtex4-1}
\usepackage{amsfonts}
\usepackage{amsmath}
\usepackage{amssymb}
\usepackage{graphicx}
\usepackage{appendix}

\usepackage[bookmarks=true,bookmarksopen,bookmarksnumbered,
			colorlinks,linkcolor=blue,citecolor=blue,
            breaklinks=true]{hyperref}

\newcommand{\grep}{\gamma_{\text{rep}}}

\begin{document}

\title{Pulsed electron spin resonance spectroscopy in the Purcell regime} 

\author{V.~Ranjan}
\email[]{vishal.ranjan@cea.fr}

\affiliation{Quantronics group, SPEC, CEA, CNRS, Universit\'e Paris-Saclay, CEA Saclay 91191 Gif-sur-Yvette Cedex, France}

\author{S.~Probst}
\affiliation{Quantronics group, SPEC, CEA, CNRS, Universit\'e Paris-Saclay, CEA Saclay 91191 Gif-sur-Yvette Cedex, France}

\author{B.~Albanese}
\affiliation{Quantronics group, SPEC, CEA, CNRS, Universit\'e Paris-Saclay, CEA Saclay 91191 Gif-sur-Yvette Cedex, France}

\author{A.~Doll}
\affiliation{Laboratoire Nanomagn\'etisme et Oxydes, SPEC, CEA, CNRS, Universit\'e Paris-Saclay, CEA Saclay 91191 Gif-sur-Yvette Cedex, France}

\author{O.~Jacquot}
\affiliation{Quantronics group, SPEC, CEA, CNRS, Universit\'e Paris-Saclay, CEA Saclay 91191 Gif-sur-Yvette Cedex, France}

\author{E.~Flurin}
\affiliation{Quantronics group, SPEC, CEA, CNRS, Universit\'e Paris-Saclay, CEA Saclay 91191 Gif-sur-Yvette Cedex, France}

\author{R.~Heeres}
\affiliation{Quantronics group, SPEC, CEA, CNRS, Universit\'e Paris-Saclay, CEA Saclay 91191 Gif-sur-Yvette Cedex, France}

\author{D.~Vion}
\affiliation{Quantronics group, SPEC, CEA, CNRS, Universit\'e Paris-Saclay, CEA Saclay 91191 Gif-sur-Yvette Cedex, France}

\author{D.~Esteve}
\affiliation{Quantronics group, SPEC, CEA, CNRS, Universit\'e Paris-Saclay, CEA Saclay 91191 Gif-sur-Yvette Cedex, France}

\author{J.~J.~L.~Morton}
\affiliation{London Centre for Nanotechnology, University College London, London WC1H 0AH, United Kingdom}

\author{P.~Bertet}
\affiliation{Quantronics group, SPEC, CEA, CNRS, Universit\'e Paris-Saclay, CEA Saclay 91191 Gif-sur-Yvette Cedex, France}

\date{\today}

\begin{abstract}
When spin relaxation is governed by spontaneous emission of a photon into the resonator used for signal detection (the Purcell effect), the relaxation time $T_1$ depends on the spin-resonator frequency detuning $\delta$ and coupling constant $g$. We analyze the consequences of this unusual dependence for the amplitude and temporal shape of a spin-echo in a number of different experimental situations. When the coupling $g$ is  distributed inhomogeneously, we find that the effective spin-echo relaxation time measured in a saturation recovery sequence strongly depends on the parameters of the detection echo. When the spin linewidth is larger than the resonator bandwidth, the Fourier components of the echo relax with different characteristic times, which implies that the temporal shape of the echo becomes dependent on the repetition time of the experiment. We provide experimental evidence of these effects with an ensemble of donor spins in silicon at millikelvin temperatures measured by a superconducting micro-resonator. 
\end{abstract}


\maketitle

\section{introduction}
Pulsed magnetic resonance spectroscopy proceeds by applying sequences of control pulses to an ensemble of electron or nuclear spins via an electromagnetic resonator of frequency $\omega_0$ (at microwave or radio frequency, respectively). Driven by these pulses, the spins undergo rotations on the Bloch sphere. Spins with identical Larmor and Rabi frequencies (forming a {\it spin packet}) follow the same trajectory. A prominent pulse sequence is the Hahn echo: a first pulse imprints its phase coherence among all spin packets, which quickly vanishes due to the inhomogeneity in Larmor precession frequency. Coherence is restored by a second control pulse applied after a delay $\tau$, which imposes a $\pi$ phase shift to the spin packets leading to their collective rephasing after another delay $\tau$. This causes the buildup of a macroscopic oscillating magnetization and the emission of a pulse known as the spin-echo, whose amplitude, shape, and time-dependence bear the desired information on spin characteristics and environment~\cite{schweiger_principles_2001}.

The maximum spin-echo amplitude is governed by the equilibrium spin longitudinal magnetization $S_{z0}$. After each echo sequence, the spins are strongly out of equilibrium, so that before the next sequence can be started, a waiting time is needed for the longitudinal spin polarization $S_z$ to relax back towards $ S_{z0}$ by energy exchange between each spin and its environment in a characteristic time $T_1$.

In solids, the dominant energy exchange processes are usually spin-phonon or spin-spin interactions. For isotropic systems, $T_1$ depends only on global sample properties (temperature, concentration in magnetic species, ...); as a result, all spin packets contributing to the echo emission relax in the same way~\cite{tyryshkin_electron_2012}. In anisotropic systems, correlations may exist between $T_1$ and the Larmor frequency of the spins~\cite{schweiger_principles_2001,du_temperature_1995}. Spin and spectral diffusion, as well as polarization transfer mechanisms, may also play a role and lead to non-exponential $S_z$ relaxation~\cite{koptyug_inversion-recovery_1996,doll_adiabatic_2013}.

Spins can also relax to equilibrium by exchanging energy with the radiation field. In free space, radiative relaxation is negligibly slow ($\sim 10^{12}$\,s for an electron spin at $9$\,GHz); it can however be considerably accelerated by the coupling to the detection resonator\cite{purcell_purcell_1946} because of the spatial and spectral field confinement brought by the latter. In special conditions, radiative relaxation can even become the dominant spin relaxation process for return to equilibrium. This Purcell regime was reached recently for an ensemble of electron spins coupled to a superconducting micro-resonator at millikelvin temperatures\cite{bienfait_controlling_2016,eichler_electron_2017}. The strength of radiative effects is characterized by the spin-resonator coupling constant $g$, defined as half the Rabi frequency a spin would have if it was driven by a resonator field with an average energy of $1$ photon, whose amplitude we denote as $\delta B_1$. The Purcell relaxation time is then given by

\begin{equation}
    \label{eq:Purcell}
        T_1 = \frac{\kappa}{4 g^2} \left[ 1 + \left( \frac{2\delta}{\kappa} \right)^2 \right],
\end{equation} 
where $\kappa = \omega_0 / Q$ is the resonator energy damping rate, $Q$ being its quality factor, and $\delta$ is the spin-resonator frequency detuning. At resonance, $T_1= \kappa / 4g^2$, which shows that Purcell relaxation is enhanced for resonators with high quality factor and small mode volume. We stress that Purcell relaxation should be distinguished from radiation damping, which is a conservative deterministic radiative feedback mechanism induced by the coherent collective  precession of the spins, and cannot cause their return to thermal equilibrium from an arbitrary (possibly unpolarized) spin state ~\cite{bloembergen_radiation_1954,augustine_transient_2002,bienfait_controlling_2016,mao_radiation_1994}.

One noticeable aspect of Eq.~\ref{eq:Purcell} is that Purcell relaxation is a resonant phenomenon, with a strong frequency dependence over a scale given by the resonator linewidth $\kappa$, as observed experimentally\cite{bienfait_controlling_2016,eichler_electron_2017}. Another consequence of Eq.~\ref{eq:Purcell} is that $T_1$ depends on the position $r$ of a given spin within the resonator mode, since $g$ is proportional to the field mode amplitude $\delta B_1(r)$. Therefore, in the Purcell regime, spin packets with different detuning and Rabi frequency also have a different relaxation time $T_1$, an unusual situation in magnetic resonance. Because a spin echo is the sum of the contribution of all spin packets, spin-echo relaxation is expected to display a complex behavior, particularly when either the Larmor frequency or the Rabi frequency are inhomogeneously distributed. 

It is the purpose of this article to analyze the implications of these correlations between relaxation time, detuning and Rabi frequencies on the shape and time-dependence of the spin-echo in the Purcell regime. We first provide a simplified model that yields analytical results for the spin-echo amplitude and shape in the Purcell regime, and enables us to identify qualitatively novel effects. When the Rabi frequency is inhomogeneously distributed, the spin-echo amplitude is found to come back to equilibrium with an approximately exponential temporal dependence, but with a time constant that depends on the control pulse amplitude. When the Larmor frequency is inhomogeneously distributed over a frequency range broader than the resonator linewidth, the various Fourier components of a spin echo relax with different time constants, which also implies that the echo shape becomes dependent on the waiting time between consecutive sequences. The third section describes the experimental setup and samples used to test these effects, and the fourth section presents the measurements, their qualitative agreement with our simplified model, and their quantitative agreement with the simulation of the Bloch equations including explicitly the Purcell decay contribution.

\section{Spin-echo in the Purcell regime : a simple model}

In this section we analyze a simple model which exemplifies the consequences of Purcell relaxation by considering analytically-tractable limiting cases. 

\subsection{System description and equations of motion}

\begin{figure}
	\center
	\includegraphics[width=\columnwidth]{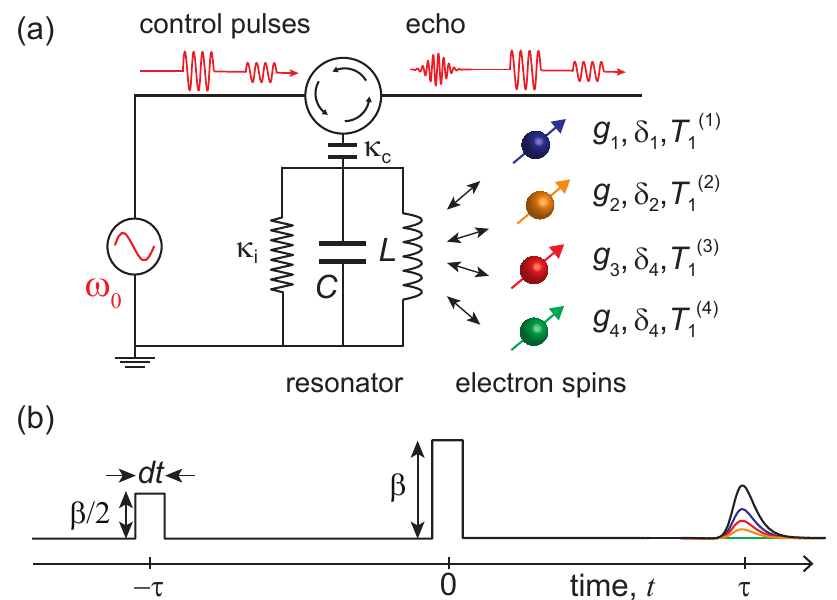}
	\caption{ A Sketch of electron-spin resonance spectroscopy in the Purcell regime. (a) An ensemble of spins is coupled to a LC resonator of frequency $\omega_0$ and internal energy decay rate $\kappa_i$. The resonator is coupled (energy decay rate $\kappa_c$) to a drive and measurement line. Each spin can have a different coupling  $g_j$ and detuning $\delta_j$ with the resonator. Spins that have same $g$ and $\delta$, and hence the same Rabi and Larmor frequencies, form a packet and have the same Purcell relaxation time $T_1$. (b) A typical Hahn echo sequence with rectangular pulses. Final echo contains contributions from different spin packets.  The relative amplitude of contributions depends on experimental parameters such as the pulse amplitude $\beta$ and the repetition rate.}
	\label{fig:ESRsetup}
\end{figure}

The system to be modelled is shown in Fig.~\ref{fig:ESRsetup}. An ensemble of spins $S=1/2$ interact with the microwave field inside the resonator used for inductive detection, which is capacitively coupled to a measurement line. The resonator is characterized by its frequency $\omega_0$, and total energy damping rate $\kappa$. Energy is lost by leakage into the measurement line (with a rate $\kappa_c$) and internal losses (rate $\kappa_i$), with $\kappa = \kappa_c + \kappa_i$. Control pulses at $\omega_0$ are sent to the resonator input through the measurement line, into which the subsequent spin echo signals are emitted then routed via a circulator towards the detection chain.

Each of the $j=1,..,N$ spins is characterized by its Larmor frequency $\omega_j$ (or equivalently the spin-cavity detuning $\delta_j = \omega_j - \omega_0$) and its coupling to the resonator field $g_j$. In the weak spin-resonator coupling limit $g_j \ll \kappa$, the spin-resonator quantum correlations can be neglected. The dynamics is well described by equations involving only the expectation value of the spin and resonator field operators \cite{ansel_optimal_2018}, written in the frame rotating at $\omega_0$ as

\begin{eqnarray}\label{eq:Bloch}
\begin{cases}
\dot{S}_x^{(j)} = - \delta_j S_y^{(j)} + 2 g_j Y S_z^{(j)} - S_x^{(j)} / T_2^{(j)}  \\
\dot{S}_y^{(j)} = \delta_j S_x^{(j)} - 2 g_j X S_z^{(j)} -  S_y^{(j)} / T_2^{(j)}  \\
\dot{S}_z^{(j)} =  2 g_j X S_y^{(j)} - 2 g_j Y S_x^{(j)} - (S_z^{(j)} - S_{z0})/ T_1^{(j)}.
\end{cases}
\end{eqnarray}

Here, $S_{x,y,z}^{(j)}$ is the expectation value of the corresponding dimensionless spin $j$ operator (with $S_{x,y,z}^{(j)} = \sigma_{x,y,z}/2$, $\sigma_{x,y,z}$ being the Pauli operators) and $X,Y$ is the expectation value of the intra-resonator field quadrature operators expressed in dimensionless units \cite{ansel_optimal_2018}. Here we use the convention that $X=(a + a^\dagger)/2$ and $Y=i (a - a^\dagger)/2$, $a$ (resp. $a^\dagger$) being the resonator annihilation (resp. creation) operator, which differs from the one used in Refs.~\onlinecite{ansel_optimal_2018,probst_shaped_2019} by a factor $2$. Phase coherence of spin $j$ relaxes in a time $T_2^{(j)}$, and $S_z^{(j)}$ relaxes towards the thermal equilibrium polarization $S_{z0} = \tanh (\hbar \omega_0 / 2 k T_s)$ in a time $T_1^{(j)}$. Because we are specifically interested in the impact of Purcell relaxation on spin-echo signals, we will in the following consider that the only decoherence mechanism is the Purcell relaxation, implying that $T_1^{(j)} = \frac{\kappa}{4 g_j^2} [1 + (\frac{2\delta_j}{\kappa})^2]$ and $T_2^{(j)} = 2 T_1^{(j)}$. 

Note that these equations are identical to the usual Bloch equations, with the Rabi frequency given by $2 g_j \alpha(t)$, and $\alpha(t) \equiv \sqrt{X(t)^2 + Y(t)^2}$. Purcell relaxation appears as a $T_1$ mechanism, and its particular features arise from the dependence of $T_1$ on the properties of a spin packet. 

Using standard input-output theory\cite{gardiner_input_1985}, the resonator field quadratures in the rotating frame at $\omega_0$ obey Eqs.

\begin{eqnarray}\label{eq:CavField}
\begin{cases}
\dot{X}(t) = \sqrt{\kappa_c} \beta_X(t) - \frac{\kappa}{2} X(t) -  \sum_{j=1}^{N} g_j S_y^{(j)}  \\
\dot{Y}(t) = \sqrt{\kappa_c} \beta_Y(t) - \frac{\kappa}{2} Y(t) +\sum_{j=1}^{N} g_j S_x^{(j)},
\end{cases}
\end{eqnarray}
where $\beta_{X,Y}(t)$ are the input control field quadratures. We also need to compute the field leaking out of the cavity, as it contains the spin free-induction-decay and echo signals. The output field quadratures ${X_\text{out}(t),Y_\text{out}(t)}$ are given by the input-output relations

\begin{eqnarray}\label{eq:out}
\begin{cases}
X_\text{out}(t) = \sqrt{\kappa_c} X(t) - \beta_X(t)  \\
Y_\text{out}(t) = \sqrt{\kappa_c} Y(t) - \beta_Y(t).
\end{cases}
\end{eqnarray}

The combination of Eqs.~\ref{eq:Purcell} to \ref{eq:out}
is the theoretical framework that describes EPR spectroscopy in the Purcell regime. All the specific characteristics of the spin ensemble are provided by the distributions of Larmor frequency $\rho_\delta(\delta)$ and coupling constant $\rho_g(g)$, normalized such that $\int dg \rho_g (g) = \int d\delta \rho_\delta (\delta) = 1$. We also define $\Gamma$ as the standard deviation of $\rho_\delta$, thus corresponding to the spin ensemble inhomogeneous linewidth. 

Although this is not immediately apparent, we note that radiation damping is automatically included in the above equations, since they treat on an equal footing the intra-resonator field and the spin operators and thus  incorporate all feedback effects of the radiation field on the spin dynamics~\cite{bloembergen_radiation_1954,ansel_optimal_2018}. This underlines the distinction between radiative damping and Purcell relaxation, which both are radiative effects but with different characteristics and impact on the spin dynamics.

Throughout this article, we make the extra simplifying hypothesis that the spin-ensemble cooperativity $C=\sum g_j^2/(\kappa \Gamma)$ verifies $C \ll 1$. In this limit, the field generated by the spins is small compared to the intra-resonator field \cite{julsgaard_quantum_2013} so that radiation damping can be entirely neglected. One can thus 1) compute the intra-resonator field $[X(t),Y(t)]$ using Eq.~\ref{eq:CavField} with the last term neglected, 2) use it to solve the spin dynamics (Eq.~\ref{eq:Bloch}), and 3) compute the output field with Eqs.~\ref{eq:CavField} and \ref{eq:out}. This is the approach that is used to simulate numerically the spin-echo signals under arbitrary control pulse sequences, and in the next sections to derive approximate expressions for the echo amplitude. More details on the simulations can be found in the Appendix. 
\subsection{Hahn echo amplitude}

We now derive an approximate analytical expression for the amplitude of a Hahn echo, based on Eqs.~\ref{eq:Purcell} to \ref{eq:out} and a number of simplifying assumptions.

To simplify the discussion, we suppose that the control pulses generate a quasi-instantaneous intra-resonator field with a simple rectangular time-dependence of duration $dt$. This can be achieved if $dt \gg \kappa^{-1}$, or by using shaped pulses that compensate for the finite resonator response time \cite{tabuchi_total_2010,ansel_optimal_2018,probst_shaped_2019}. We thus consider input pulses on the $X$ quadrature, with an amplitude $\beta_X \equiv \beta$ during $dt$ while $\beta_Y = 0$, related to the input power $P_\text{in}$ as $\beta = \sqrt{P_\text{in}/\hbar \omega_0}$. The corresponding intra-resonator field amplitude is $\alpha = 2 \sqrt{\kappa_c} \beta / \kappa$.

Under each control pulse, spin $j$ undergoes a Rabi rotation~\cite{schweiger_principles_2001} with a frequency $\sqrt{(2 g_j \alpha)^2 + \delta_j^2}$. We assume that $\Gamma \ll 2 g_j \alpha$ for all spins so that the dependence of the Rabi frequency on $\delta_j$ can be neglected. Also, we assume that $\rho_\delta$ is symmetric and is excited in its centre.

The Hahn-echo pulse sequence is shown in Fig.~\ref{fig:ESRsetup}. A first pulse of amplitude $\beta / 2$ and duration $dt$ is followed by a waiting time $\tau$, then by a second pulse of amplitude $\beta$ and same duration, and by a second waiting time $t$. An echo is formed around $t = \tau$ because of the refocusing of the spin packets (we assume $dt \ll \tau$). We also use an equivalent Hahn-echo sequence where the two pulses have the same amplitude $\beta$, but the first pulse duration is $dt/2$. The spin-echo originates solely from the $y$ component of the magnetization, and the contribution of spin $j$ can be shown~\cite{bloom_nuclear_1955,schweiger_principles_2001} to be

\begin{eqnarray}
    S_y^{(j)} (\tau + t)&  = & - S_z^{(j)}  \sin ^3 (2 \alpha g_j dt) \cos \delta_j (t - \tau).
\end{eqnarray}

\noindent where $S_z^{(j)}$ is spin $j$ longitudinal polarization at the time of the first control pulse. The latter is not necessarily equal to $S_{z0} $ because the waiting time since the previous pulse sequence may not be sufficiently long. Importantly, because of Purcell relaxation, $S_z^{(j)}$ depends on $g_j$ and $\delta_j$, which leads to novel effects as shown below. Note that we have, however, neglected the impact of spin relaxation during the Hahn echo sequence, because in most relevant cases $\tau \ll T_1^{(j)}$.

To obtain simple expressions for the echo amplitude, we also consider that the resonator field dynamics adjusts adiabatically to the spin operators, as would be the case in the limit of low resonator $Q$. Equations \ref{eq:CavField} and \ref{eq:out} then yield

\begin{multline}
    \label{eq:Echoamp}
    X_\text{out}(\tau + t) = - 2 \frac{\sqrt{\kappa_c}}{\kappa} \sum_j g_j S_y^{(j)} (\tau + t) \\
    =  - 2 \frac{\sqrt{\kappa_c}}{\kappa} \int \int g \rho_g(g) \rho_\delta (\delta)  S_y(g,\delta,t) d\delta dg, 
\end{multline}
where we have taken the continuous limit in the last expression and defined $S_y(g,\delta,t)$ as being equal to $S_y^{(j)} (\tau + t)$ for a spin $j$ having a coupling $g_j=g$ and detuning $\delta_j= \delta$. Note also that we have implicitly assumed that there is no correlation between the frequency of a given spin and its coupling constant $g$, an assumption that is not always verified \cite{pla_strain-induced_2018}.

This echo amplitude $X_\text{out} (t+\tau)$ depends in an intricate way on the pulse amplitude and on the shape of the inhomogeneous distributions $\rho_g$ and $\rho_\delta$. In the following we consider two limiting cases.
\subsection{Narrow-line case}

\begin{figure}
	\center
	\includegraphics[width=\columnwidth]{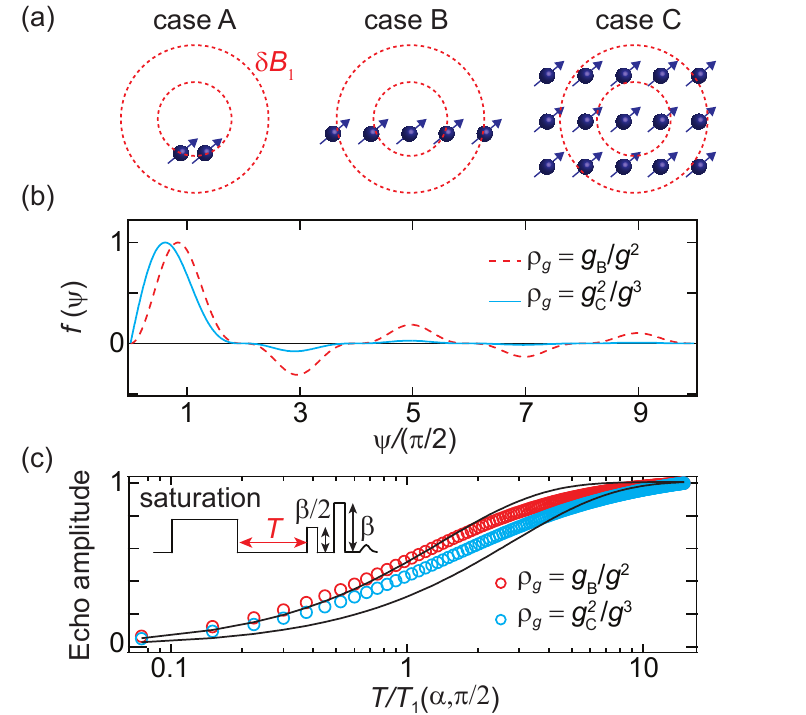}
	\caption{ Rabi angle selectivity in the narrow line case. (a) Inhomogeneous coupling of spins to the resonator for case A: point distribution, case B: thin layer distribution, case C: bulk distribution. Dashed lines represent different contours of spin-photon coupling strengths.  (b) Calculated relative echo contribution $f(\psi)$ from spin packets with different Rabi angle $\psi$. (c) Saturation recovery of the echo amplitude calculated using the model. We have taken small Rabi angles $0<\psi<\pi$. Solid black lines are exponential with decay constant $T_1~(\alpha,\psi_0)$, where $\psi_0$ is the value where $f(\psi)$ is maximum in panel (b).  }
	\label{fig:spindistribution}
\end{figure}

Let us first assume that the spin ensemble has a linewidth much narrower than the resonator, ie $\Gamma \ll \kappa$. Then, $T_1 \approx \kappa/(4 g^2)$, implying that $ S_z$ and $S_y$ do not depend any longer on $\delta$. Overall, the echo amplitude at $t=\tau$ becomes

\begin{equation}
\label{eq:EchoNarrowLine}
    X_\text{out} (2\tau) = 2 \frac{\sqrt{\kappa_c}}{\kappa} \int g  S_z  \rho_g(g) \sin^3 \big( 2 \alpha g dt \big) dg.
\end{equation}
To appreciate the impact of Purcell relaxation, we now focus on the so-called saturation recovery sequence schematically depicted in Fig.~\ref{fig:spindistribution}c, which is commonly used to measure spin relaxation time. It consists of applying a saturating pulse at time $t=0$ (so that $S_z(t=0)=0$ for all spins), followed by a Hahn-echo sequence applied after a waiting time $T$ at $t=T$. Its maximum amplitude at time $t=T+2\tau$ is denoted by $X_\text{out}(T)$ for simplicity in the following. At the beginning of the Hahn-echo sequence, 

\begin{equation}
S_z (T) = S_{z0} \left( 1 - e^{-T/T_1} \right).
\label{eq:SzSaturationRecovery}
\end{equation}

In the usual situation where spin relaxation is not Purcell-limited as is the case when it is governed by interactions with the lattice, and where the system is moreover isotropic, all spins then relax with the same time constant $T_1$. In Eq.~\ref{eq:EchoNarrowLine}, $S_z$ can be factorized ; the echo amplitude is proportional to $S_z(T)$, and it follows the exponential dependence of $S_z(T)$, which enables to measure $T_1$. The echo amplitude effectively measures the longitudinal polarization $S_z(T)$, which justifies its denomination as a {\it detection echo}; in particular, its parameters (pulse amplitude, duration, ...) have no impact on the measured $T_1$. 

In the Purcell regime, $T_1$ is different for spin-packets with different couplings $g$; the detection echo amplitude $X_\text{out}(T)$ has thus no reason to be even exponential. Because of the strong $g$-dependence of the integrand in Eq.~\ref{eq:EchoNarrowLine} however, an approximate exponential dependence is nevertheless recovered in a number of cases, but with an effective relaxation time that now depends on the parameters of the detection echo.

We now make this reasoning explicit by considering three examples of coupling constant distributions $\rho_g$ shown in Fig.~\ref{fig:spindistribution}. The spins may be located in an area where $B_1$ is very homogeneous (case A), or on the contrary very inhomogeneous (cases B and C). To consider cases that correspond to situations encountered in recent experiments, we assume that $B_1$ is generated by a narrow wire deposited at the surface of a sample containing the spins. The spins may be distributed within a thin layer just below the sample surface (case B), or homogeneously in the bulk (case C). Assuming that $\delta B_1(r) = \mu_0 \delta i / (2 \pi r) $, $r$ being the spin-wire distance and $\delta i$ the quantum fluctuations of the ac current in the resonator, it is straightforward to see that $\rho_g(g) = \delta (g-g_A)$ in case A, $\rho(g) = g_B/ g^2$ in case B, and $\rho(g) = g_C^2/g^{3}$ in case C. 

For completeness, it is interesting to note that the situation may be complicated by the presence of correlations between the spin Larmor and Rabi frequency. This was the case in the experiments reported in Ref.~\onlinecite{bienfait_reaching_2015,bienfait_controlling_2016}. There, silicon donor spins were confined to a thin layer ($100$\,nm) below the surface of a silicon sample, on top of which a thin-film resonator with a $5 \mu \mathrm{m}$-wide inductance was deposited. Because of the mechanical strain exerted by the thin metallic film on the silicon substrate \cite{pla_strain-induced_2018} due to differential thermal contractions between the metal and the silicon, the spin hyperfine interaction becomes correlated with the lateral position relative to the wire, which also happens to be approximately correlated to $B_1$. As a result, the Larmor and Rabi frequencies of the spins are correlated, and by properly choosing the biasing field $B_0$ the system is much better described by case A than case B. This approximate correlation is valid only when the wire transverse dimensions are large compared to the spin layer thickness, and was indeed no longer found in Ref.~\onlinecite{probst_inductive-detection_2017} where the wire width was decreased to $500$\,nm. 

We will now compute the signal expected from a saturation recovery sequence in the three afore-mentioned cases, by inserting Eq.~\ref{eq:SzSaturationRecovery} into Eq.~\ref{eq:EchoNarrowLine}. 

In case A, we obtain straightforwardly 

\begin{equation}
\label{eq:CaseA}
    X_\text{out} (T) = 2 \frac{\sqrt{\kappa_c}}{\kappa} g_A \sin^3(2 \alpha g_A dt) S_{z0} \big(1 - \exp^{-\frac{T}{T_1(g_A)}} \big).
\end{equation}

Because the coupling constant has a well-defined value $g_A$, the spin relaxation time is identical for all measured spins and we are in the same situation as in usual magnetic resonance, where the detection echo amplitude relaxes exponentially with the Purcell relaxation time $T_1(g_A)$, independently of $\alpha$. 

To deal with cases B and C, it is useful to introduce the Rabi rotation angle $\psi = 2 \alpha g dt$. The $T$ dependent echo amplitude can then be written as

\begin{equation}
\label{eq:EchoNarrowLineVsT}
    X_\text{out} (T) = 2 \frac{\sqrt{\kappa_c}}{\kappa} S_{z0} \int \big[1 - \exp^{-\frac{T}{T_1(\alpha,\psi)}} \big] f(\psi) d\psi,
\end{equation}

\noindent with $T_1(\alpha,\psi) = \frac{\kappa \alpha^2 dt^2}{\psi^2}$ and $f(\psi) = \psi \rho_\psi \sin^3 \psi / (2 \alpha dt)$. The function $f(\psi)$ indicates the relative contribution of spin packets to the total echo signal as a function of their Rabi angle $\psi$. As shown in Fig.~\ref{fig:spindistribution} for cases B and C, it displays a maximum at a value $\psi_0$ close to $\pi / 2$. In a crude approximation, $f(\psi) \sim \delta (\psi - \psi_0)$, so that 

\begin{equation}
\label{eq:SingleExp}
    X_\text{out} (T) = 2 \frac{\sqrt{\kappa_c}}{\kappa} S_{z0}\big(1 - \exp^{-\frac{T}{T_1(\alpha,\psi_0)}} \big).
\end{equation}

\noindent Despite the broad coupling constant distribution, one thus recovers an approximate exponential dependence of the detection echo amplitude on the waiting time $T$. However, the effective measured relaxation time scales like $\alpha^2$, the square of the amplitude of the pulse used in the detection echo sequence. The physical interpretation is straightforward : the detection echo signal mostly originates from the contribution of spins that undergo first a $\pi/2$ pulse and then a $\pi$ pulse during the refocusing step. When the pulse amplitude is varied, this amounts to selecting spins with different coupling constants $g$, and therefore different relaxation times. 

To verify the validity of approximating $f(\psi)$ by a $\delta$ function, we compute Eq.~\ref{eq:EchoNarrowLineVsT} numerically for cases B and C. The results are shown in Fig.~\ref{fig:spindistribution}(c) for the decay of the spin-echo amplitude $A(T)$ (dots), compared to the single-exponential approximation of Eq.~\ref{eq:SingleExp} (solid lines, taking into account the different values of $\psi_0$ for cases B and C). The qualitative agreement demonstrates that Eq.~\ref{eq:SingleExp} correctly captures the impact of Purcell relaxation on the effective relaxation time measured in a saturation-recovery sequence.

\subsection{Narrow-coupling case}

We now consider the case where the coupling constant is single-valued (corresponding to case A in the previous section) $\rho_g(g) = \delta (g-g_A)$. The pulse amplitude is chosen such that $2 g_A \alpha dt = \pi/2$, so that the control pulses implement the ideal Hahn-echo sequence. The Larmor frequency on the other hand is broadly distributed, with $\Gamma \gg \kappa$. 

In that limit, the Hahn echo amplitude Eq.~\ref{eq:Echoamp} becomes 

\begin{equation}
\label{eq:echoshape}
    X_\text{out}(\tau + t) = 2 \frac{\sqrt{\kappa_c}}{\kappa} \int S_z \rho_\delta \cos \delta (t-\tau) d \delta.
\end{equation}
Consider now that the echo sequences are repeated multiple times, with a waiting time $T$ in-between two consecutive sequences, and let us assume that the spins are fully un-polarized at the immediate end of a sequence, $S_z = 0$. Then at the beginning of each echo, the longitudinal polarization $S_z$ is given by Eq.~\ref{eq:SzSaturationRecovery}. 

In a usual situation where $T_1$ is not correlated with the spin Larmor frequency, $S_z$ can be factored out of the integral in Eq.~\ref{eq:echoshape}, and the echo temporal shape is simply given by the Fourier transform of the Larmor frequency distribution $\rho_\delta(\delta)$.

In the Purcell regime however, we get that

\begin{equation}
    S_z = S_{z0} \big(1 - e^{-T/T_1(\delta)}\big),
\end{equation}
so that 

\begin{multline}
\label{eq:FFTTheory}
     X_\text{out}(\tau + t)  =  2 \frac{\sqrt{\kappa_c}}{\kappa} S_{z0}~\times  \\
     \int \big(1 - e^{- T/T_1(\delta)}\big) \rho_\delta(\delta) \cos \delta (t-\tau) d\delta,    
\end{multline}
whose Fourier transform is

\begin{equation}
    \tilde{X}_\text{out}(\delta,T)= 2 \frac{\sqrt{\kappa_c}}{\kappa} S_{z0} (1 - e^{-T/T_1(\delta)}\big) \rho_\delta(\delta).
\end{equation}

The Fourier components $\tilde{X}_\text{out}(\delta)$ of the echo therefore relax with a time constant $T_1(\delta)$ that follows the Purcell law, and should thus increase with detuning quadratically. 

Moreover, the spin-echo temporal shape is given by the Fourier transform of $\rho_\delta(\delta) \big(1 - e^{-T/T_1(\delta)}\big)$, which now depends on the repetition time $T$. The interpretation is here again straightforward. Because of the dependence of the relaxation time on the detuning, the physical spin distribution $\rho_\delta(\delta)$ is effectively renormalized by the factor $\big(1 - e^{-T/T_1(\delta)}\big)$. At short times $4 g_A^2 T \ll \kappa$, $\big(1 - e^{-T/T_1(\delta)}\big) \simeq  T / T_1 = 4 g^2 T / \kappa \times 1/ (1 + 4 \delta ^2 / \kappa^2)$, implying that the effective spin distribution is given by the resonator filter function. In the long time limit $4 g_A^2 T \gg \kappa$, this effective distribution is closer to the physical spin distribution function $\rho_\delta(\delta)$. Note that this Purcell-filtering of the spin distribution should not be mistaken for the electromagnetic filtering of the spin-echo signal by the cavity, which will always be there in a real experiment but was neglected here for simplicity because of the low-Q assumption.

\section{Materials and methods}

We now describe the sample and setup used to demonstrate the effects discussed above. 
Reaching the Purcell regime requires spins with long intrinsic spin relaxation times, and resonators with a small mode volume and high quality factor. We use the electron spin of donors in silicon, which have been shown to reach the Purcell regime when coupled to superconducting micro-resonators at millikelvin temperatures \cite{bienfait_controlling_2016}. 

\begin{figure}[htbp]
	\center
	\includegraphics[width=\columnwidth]{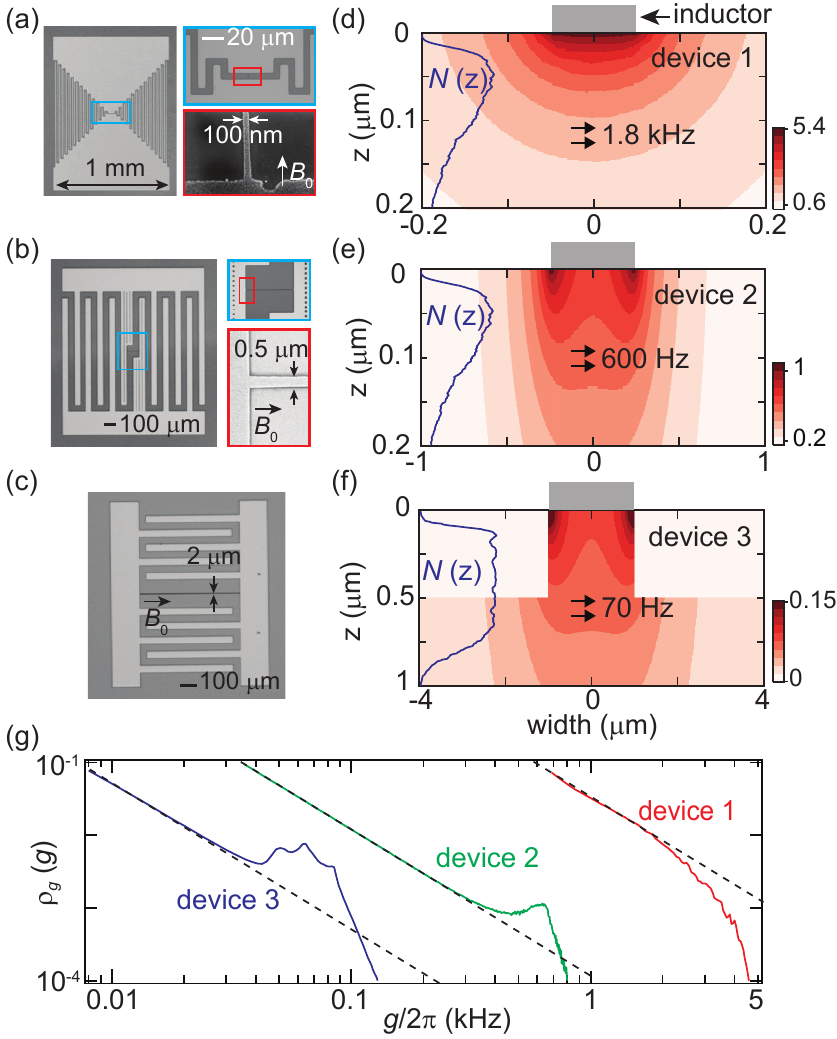}
	\caption{ Device details. (a-c) Optical photographs of the three $LC$ resonators, and scanning electron microscopy images of the sub-$\mu$m wide inductors for devices 1 and 2. The direction of the static magnetic field $B_0$ is also shown. (d-f)  Calculated coupling strength $g/2\pi$~(in kHz) distribution (color contour plots) for the 3 devices. The Bi dopants are implanted homogeneously over the whole sample area with an implantation profile $N(z)$ (left trace); the spin signal originates dominantly from spins located directly below the inductor. In panel (f), white rectangular areas on two sides of the inductor represent regions that have been etched during fabrication. (g) Log-log plot of calculated coupling distribution $\rho_g(g)$. Dashed lines represent $1/g^2$ dependence.}
	\label{fig:setup}
\end{figure}

In this work we present data from three different devices. Each device is a silicon sample that was implanted with bismuth atoms close to its surface, and on which a discrete-element superconducting LC resonator was patterned. The resonators consist of an interdigitated capacitance shunted by a micron- or sub-micron-scale wire which plays the role of the inductance [Fig.~\ref{fig:setup}(a-c)]. The devices are mounted in a copper sample holder, and coupled capacitively to a microwave antenna which determines the coupling rate $\kappa_c$. In devices 1 and 2 [Fig.~\ref{fig:setup}(d-e)], the implantation depth is $\sim 100$~nm, with a peak concentration of $8\times 10^{16}~\mathrm{cm}^{-3}$, the silicon sample is isotopically enriched in $^{28}\mathrm{Si}$, and the resonator is made in aluminum. In device 3 [Fig.~\ref{fig:setup}(f)], the implantation depth is $\sim 1~ \mu \mathrm{m}$, with a smaller peak concentration of $10^{16}~\mathrm{cm}^{-3}$, the silicon is of natural isotopic abundance, and the resonator is made of niobium. Of particular importance for this work is the geometry of the resonator inductance, which strongly impacts the coupling constant distribution $\rho_g(g)$. It is $100$\,nm wide and $10~\mu \mathrm{m}$ long in device 1, $500$\,nm wide and $100~\mu \mathrm{m}$ long in device 2, and $2~\mu \mathrm{m}$ wide and $700~\mu \mathrm{m}$ long in device 3. The characteristics of the devices are summarized in Table~\ref{table1}.

\begin{table}
\caption{Device parameters}
\label{table1}
\begin{center}
  \begin{tabular}{ | c | c | c | c | }
    \hline
     & Device 1 & Device 2 & Device 3 \\ 
     \hline
    Substrate  & $^{28}$Si & $^{28}$Si & natural Si \\ 
    \hline
    $^{29}$Si in substrate  ($\%$) & 0.05 & 0.05 & 4.7 \\ 
    \hline
    $N_\text{max}$ (cm$^{-3}$) & 8 $\times 10^{16}$ & 8 $\times 10^{16}$ & $10^{16}$\\ \hline
    Implantation depth $(\mu$m) & $\sim~0.1$ & $\sim$ 0.1 & $\sim$~1\\ \hline
    Resonator material & Al & Al & Nb \\ \hline
    Inductor width ($\mu$m) & 0.1 & 0.5 & 2 \\ \hline
    Inductor length ($\mu$m) & 10 & 100 & 700 \\ \hline  
    $\omega_0/2\pi$ (GHz) & 7.25 & 7.25 & 7.41\\ \hline
    Total $Q$  & $2 \times 10^4$ & $6 \times 10^4$ & $ 10^4$ \\ \hline
    $Z_0~(\Omega)$ & 15 & 30 & 40\\ \hline
    $B_0$ (mT) & 3.7 & 3.7 & 62.5 \\ \hline
  \end{tabular}
\end{center}
\end{table}

Bismuth atoms implanted in the silicon form four covalent bonds to the silicon lattice, while the fifth valence electron is trapped at low temperatures by the hydrogenic potential formed by the ionized bismuth atom; its spin gives rise to the ESR signal~\cite{feher_electron_1959}. Due to the hyperfine interaction between the unpaired electron spin and the bismuth nuclear spin, allowed ESR transitions are found close to $\sim 7.38~$GHz at small magnetic fields (here, $B_0 \sim 5$\,mT). More details on the spin Hamiltonian and ESR transitions can be found in Ref.~\onlinecite{george_electron_2010}. In our devices, bismuth donor spins experience large strain when cooled to low temperature because of the differential thermal expansion of Al and Si~\cite{mansir_linear_2018,pla_strain-induced_2018}, which leads to ESR lines much broader than both $\kappa$ and the Rabi frequency $2 g \alpha$ in our experiments. We model this by a constant distribution $\rho_\delta$.

The coupling constant distribution $\rho_g(g)$ is computed by first estimating the rms current fluctuations in the inductance $\delta i_0 = \omega_0 \sqrt{\hbar / 2 Z_0}$, where the $LC$ resonator impedance $Z_0 = \sqrt{L/C}$ is extracted from electromagnetic simulations. This yields the position dependence of the rms magnetic field fluctuations $\delta B_1 (r)$ and of the coupling constant $g(r) = \gamma_e |\langle 0 |S_x | 1 \rangle | \delta B_1 (r)$ \cite{haikka_proposal_2017} [see Fig.~\ref{fig:setup}(b-d)]. Combined with the implantation profile, we then estimate the coupling constant distribution $\rho_g(g)$ for all three devices [solid lines in Fig.~\ref{fig:setup}(g)]. At small $g$, $\rho_g(g)$ scales like $g^{-2}$ in all devices, as expected from the analysis of section IIC for case $B$. This dependence breaks down for spins that are close to the wire (and have therefore the largest coupling to the resonator), and become sensitive to its transverse dimension, which leads to a cutoff in $\rho_g(g)$. This cutoff lies at about $4$\,kHz, $0.8$\,kHz, and $0.1$\,kHz for devices 1 to 3. Because the inductor wire is larger than the spin implantation depth in devices 2 and 3, $\rho_g$ features a shoulder at about $0.6$\,kHz and $0.06$\,kHz repsectively, due to the significant density of spins lying right below the wire and therefore seeing a more uniform $B_1$ field ; in a sense, these devices are intermediate between cases $A$ and $B$. 

\begin{figure}[htbp]
	\center
	\includegraphics[width=\columnwidth]{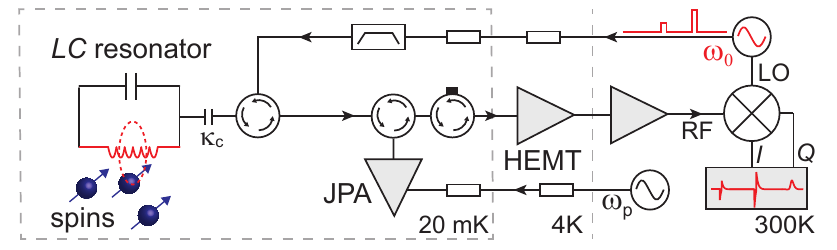}
	\caption{Schematic of the measurement setup at different temperature stages of our ESR spectrometer.  }
	\label{fig:Measurementsetup}
\end{figure}

The samples are cooled to $20~$mK in a dilution refrigerator, and measured in a setup described schematically in Fig.~\,\ref{fig:Measurementsetup}. Control pulses are sent at $\omega_0$ through an input line that incorporates low-temperature attenuation, and the reflected pulses together with the spin echo signal are routed by a circulator towards a parametric amplifier at $10$\,mK (either of the JPA~\cite{zhou_high-gain_2014} or the JTWPA \cite{macklin_nearquantum-limited_2015} type). After further amplification at $4$\,K by a High-Electron-Mobility Transistor amplifier and at $300$\,K, the output signal is homodyne demodulated by mixing with a local oscillator also at $\omega_0$. The quadrature carrying the spin-echo signal is selected, yielding the echo signal $X_\text{out}(t+\tau)$ used for analysis. 

All three resonators have slightly different frequencies $\omega_0$, all within $200$\,MHz of 7.37~GHz, the zero-field splitting of bismuth donors in silicon. The resonators are in the overcoupled regime ($\kappa_c \gg \kappa_i$), with total quality factors in the $10^4 - 10^5$ range. A magnetic field $B_0$ is applied parallel to the sample, along the inductor wire, and its value is chosen such that one of the bismuth donor transitions is resonant with $\omega_0$. 

\section{Experimental results}
\subsection{Pulse-amplitude-dependent spin relaxation}

We use device 1 to investigate the dependence of $T_1$ on the excitation amplitude $\beta$ predicted in section 2C when $\rho_g$ is broadly distributed. As discussed earlier, in device 1 $\rho_g$ corresponds well to case $B$ so that the analysis of Section IIC should apply.

Saturation recovery pulse sequences are applied, with a saturation pulse of duration $2$\,ms and fixed amplitude, followed after a variable delay $T$ by a detection echo. Square input pulses of duration $1~\mu \mathrm{s}$ are used, of respective amplitudes $\beta/2$ and $\beta$. The repetition rate of the sequence is $\grep = 30~$Hz. To more closely approximate the narrow-line hypothesis of Section 2C, the echo signal $X_\text{out}(t+\tau)$ is integrated over its duration $T_E$ yielding the echo area $A_e(T) = \int_{-T_E/2}^{+T_E/2} X_\text{out}(t+\tau) dt$, which is equal to the zero-detuning Fourier component and thus contains the contribution of spins at resonance with the resonator.

\begin{figure}
	\center
	\includegraphics[width=\columnwidth]{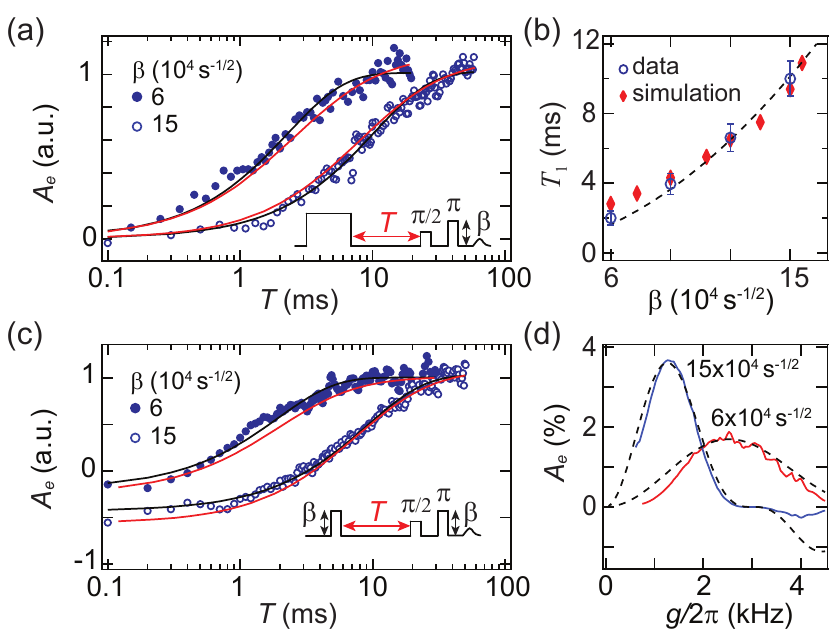}
	\caption{Spin relaxation versus excitation amplitude (device 1). (a) Relaxation of the echo area $A_e(T)$ as a function of the waiting time $T$ in a saturation recovery sequence (shown in inset), for two different values of $\beta$. Circles and red lines correspond to measurements and numerical simulations. Black lines correspond to least square fits of the data by single-exponentials. (b) Comparison of experimentally extracted $T_1$ values (open blue circles) with simulated ones (red diamonds), for more $\beta$ values. The dashed line is a quadratic fit to the data. (c) Spin relaxation measurements using inversion recovery method. (d) Calculated contribution (solid lines) of spin packets with coupling $g$ to the echo amplitude $A_e(T=30\text{ms})$, for the saturation recovery sequence and the two $\beta$ values of a). Dashed lines represent the corresponding $f(\psi)$ function introduced in the model section.}
	\label{fig:relaxation}
\end{figure}

The resulting $A_e(T)$ is shown in Fig.~\ref{fig:relaxation}(a), for two different values of $\beta$. We see that both datasets are satisfactorily fitted by exponentially decaying curves, but with different time constants $T_1$. Figure~\ref{fig:relaxation}(b) shows the measured $T_1(\beta)$, which scales like $\beta^2$ as predicted in section IIC. We also measured $T_1$ using an inversion recovery sequence [see Fig.~\ref{fig:relaxation}(c)], using an inversion pulse with the same amplitude and duration as the refocusing pulse of the detection echo. The fitted $T_1$ values are identical to the saturation-recovery ones within error bars and display the same $\beta^2$ scaling. Note that this would probably not be the case if the inversion pulse amplitude was too different from $\beta$.

\begin{figure}
	\center
	\includegraphics[width=\columnwidth]{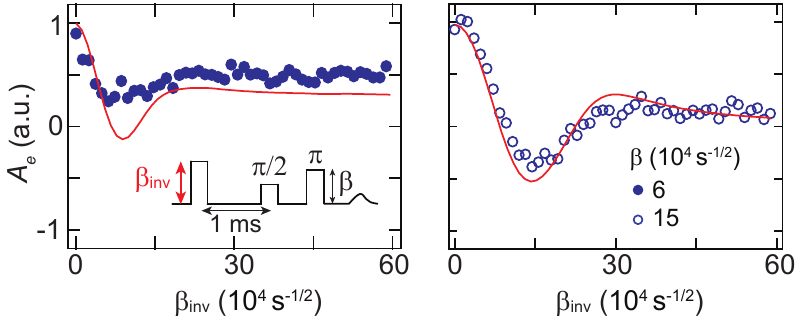}
	\caption{Rabi oscillations measured (circles) for two different values of pulse amplitude $\beta$ (see sequence in inset), using device 1. The fit to the data of simulations (solid red lines) provides a calibration of $\beta$.}
	\label{fig:Rabi}
\end{figure}

In order to compare the experiments to simulations, an absolute calibration of the input pulse amplitude $\beta$ is needed. Since attenuation and filtering along the input line cannot be known precisely enough, the calibration is performed by comparing the Rabi simulations to the dedicated Rabi pulse sequence shown in Fig.~\ref{fig:Rabi}. A first pulse of varying amplitude drives Rabi oscillations in the spin ensemble followed, after a waiting time of $1$\,ms, by detection echo. The frequency and amplitude of the resulting oscillations are compared to the simulation, which calibrates $\beta$ (see Fig.~\ref{fig:Rabi}). Using this independent calibration, we simulate the saturation recovery experiment of Fig.~\,\ref{fig:relaxation}, taking into account the estimated $\rho_g$ and $\rho_\delta$ for device A. Both the shape of the relaxation curves $A_e(T)$ and the dependence of the fitted $T_1$ on $\beta$ are well reproduced, without adjustable parameters.

To confirm the interpretation given in Section IIC, we show in Fig.~\,\ref{fig:relaxation}(d) the relative contribution to the echo of various spin packets as a function of the value of their coupling constant $g$, extracted from the simulations for the two example curves shown in Fig.~\,\ref{fig:relaxation} at a waiting time $T=30$\,ms larger than $T_1$, so that the spins are close to equilibrium. As expected, an echo obtained with a larger pulse amplitude (large $\beta$) has more contributions from spins that are more weakly coupled, compared for the echo using weaker pulses (smaller $\beta$). On the same figure we also show $f(\psi)$, and observe a satisfactory agreement with our simple model.

\subsection{Detuning dependent spin relaxation}

We investigate the detuning dependence of spin relaxation with devices 2 and 3 in which the coupling constant distribution shows a plateau (Fig.~\ref{fig:setup}), so that they approach the narrow-coupling limit described in Section IID and display well-defined Rabi oscillations allowing us to perform Rabi rotations with a well-defined angle~\cite{probst_inductive-detection_2017}.

\begin{figure}
	\center
	\includegraphics[width=\columnwidth]{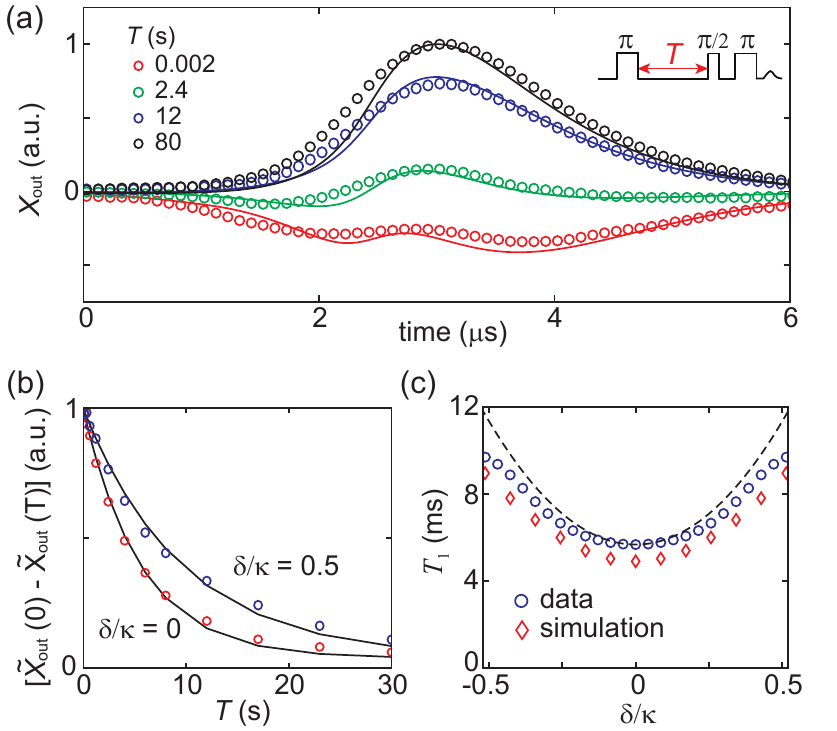}
	\caption{Spin relaxation versus detuning (device 3). (a) Echoes measured at different delays $T$ of an inversion recovery sequence (inset) applied to device $3$. Simulations shown as solid lines take into account the corresponding $\rho_g (g)$ (see Fig.\ref{fig:setup}g). (b) 
	Measured (open circles) spin relaxation $\tilde{X}_{\delta}(T)$ at two different spin detunings $\delta$, and exponential fit (lines) to the data. (c) Comparison of measured $T_1(\delta)$ (open blue circles) with simulation without adjustable parameters (open red diamonds). The dashed line shows the expected $T_1(\delta)$ dependence based on the Purcell formula Eq.~\ref{eq:Purcell} and on the measured $T_1$ at $\delta=0$.}
	\label{fig:T1_detuning}
\end{figure}

The inversion recovery sequence is first applied to device 3 (see Fig.~\ref{fig:T1_detuning}). The echo signal $X_e(t+\tau)$ is shown in Fig.~\ref{fig:T1_detuning}(a) for various values of the waiting time $T$. As expected, its phase is inverted for short values of $T$ compared to long ones. 

The Fourier transform of each curve $\tilde{X}_e(\delta)$ is then computed. Figure \ref{fig:T1_detuning}(b) shows the time dependence of two normalized Fourier components ($\delta = 0$ and $\delta = 0.5 \kappa$) and demonstrates that $\tilde{X}_{\delta=0}(T)$ relaxes faster than $\tilde{X}_{\delta=0.5\kappa}(T)$ as anticipated. $T_1(\delta)$ [see Fig.~\ref{fig:T1_detuning}(c)] is then obtained by fitting each Fourier component by an exponential decay. $T_1$ increases with $\delta$ as expected from Eq.~\ref{eq:FFTTheory}, although it does not exactly follow the Purcell law.

To understand this discrepancy, we perform numerical simulations, using the estimated $\rho_g(g)$ and a constant $\rho_\delta$ as already discussed, and without any adjustable parameter. We first compute the time traces $X_e(t+\tau)$, which we find in quantitative agreement with the data as seen in Fig.~\ref{fig:T1_detuning}(a). We then Fourier transform the simulation result and extract $T_1(\delta)$ as for the experimental data. The result [see Fig. ~\ref{fig:T1_detuning}(c)] reproduces well the dependence of $\delta$ found in the experimental values. Detailed analysis of the simulation data shows that the finite width of the $\rho_g(g)$ distribution is actually causing the discrepancy with the Purcell law: the large-$\delta$ components of the spin-echo come from spins more strongly coupled than those contributing to the $\delta=0$ component. 

\begin{figure}
	\center
	\includegraphics[width=\columnwidth]{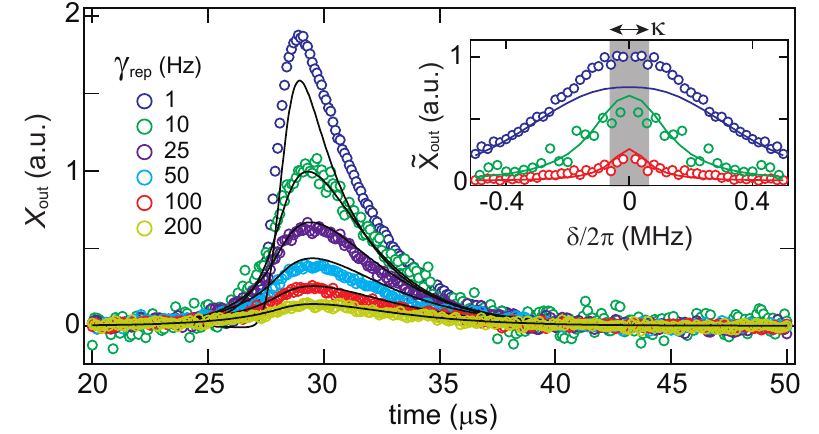}
	\caption{Influence of Hahn echo sequence repetition rate $\gamma_{\text{rep}}$ on measured (open symbols) and simulated (lines) echo shape $X_\text{out}(t)$ (device 2). Bump-shaped control pulses are used to generate $1 \mu \mathrm{s}$-long intra-resonator pulses. Inset: Fourier transform of some of the echos in the main graph (corresponding symbols, lines and colors). The resonator linewidth is indicated by a dashed area.}
\label{fig:EchoReparate}
\end{figure}

We finally test the influence of the Hahn echo sequence repetition rate $\gamma_{\text{rep}}$ on the temporal echo shape with device 2 (see Fig.~\ref{fig:EchoReparate}). In order to maximize the spin excitation bandwidth, bump-shaped excitation pulses~\cite{ansel_optimal_2018,probst_shaped_2019} are used: they are designed to make the intra-resonator field closely approximates a rectangular shape with rise and fall times much shorter than the cavity damping time $2/\kappa$, which thus brings the experiment closer to the idealized quasi-instantaneous pulse limit discussed in section II. More quantitatively, the intra-resonator field pulse length is $1~\mu \text{s}$, resulting in an excitation bandwidth of $\simeq 1$\,MHz (whereas $\kappa/2\pi = 100$\,kHz). At constant induced spin flip angle, the maximum amplitude of a bump pulse is consequently much larger than for a square pulse.

Figure \ref{fig:EchoReparate} shows that both the risetime and maximum height of the echo decrease with increasing $\gamma_{\text{rep}}$. In particular, the rise becomes faster than $2/\kappa$ at $\gamma_{\text{rep}}=1$~Hz, which confirms spin excitation outside the resonator bandwidth. The echo emission is however inevitably filtered by the cavity and the echo-shape follows a cavity ring-down with time constant $\sim 2/\kappa$. Once again, numerical simulations (see Fig.~\ref{fig:EchoReparate}) capture the changes in echo-shape and magnitude.
Similar to device 1, the pulse amplitude for device 2 ($\beta = 5 \times 10^6$~s$^{-1/2}$) is estimated from corresponding Rabi and $T_1$ measurements ($\approx 30$~ms, extracted using square pulses).

\section{Conclusion}
We have explored theoretically and experimentally the consequences of spin relaxation by the Purcell effect on the temporal shape and amplitude of spin echoes in a number of experimental situations. They result from the correlations between relaxation time of a given spin packet, its detuning to the resonator, and its spin-photon coupling constant (or equivalently its Rabi frequency). 

When the spin-resonator coupling is distributed inhomogeneously, the detection echo gets its strongest contribution from a sub-ensemble of spins that undergo rotation angles close to $\pi/2$ and $\pi$ respectively, which also determines their Purcell decay rate. As a result, the spin echo decays with an effective time constant that depends on the detection echo pulse amplitude. 

When the spin-resonator detuning distribution is broader than the resonator linewidth, the relaxation time of the spin-echo Fourier components varies quadratically with detuning, which leads to a repetition-time-dependent shape for the spin-echo. 

Our results show that in the Purcell regime, usual pulse EPR sequences should be re-analyzed in-depth; indeed, neglecting the correlations pointed out here could lead to incorrect assessments on the spin relaxation time, for instance. More generally, these qualitatively novel effects confirm that Purcell relaxation constitutes a novel regime for magnetic resonance that deserves deeper exploration on its own.

\section*{Acknowledgement}
We thank P.~S\'enat, D. Duet and J.-C. Tack for the technical support, and are grateful for fruitful discussions within the Quantronics group. We acknowledge IARPA and Lincoln Labs for providing a Josephson Traveling-Wave Parametric Amplifier used in some of the measurements. We acknowledge support of the European Research Council under the European Community's Seventh Framework Programme (FP7/2007-2013) through grant agreement No.~615767 (CIRQUSS), and of the Agence Nationale de la Rercherche under the Chaire Industrielle NASNIQ. AD acknowledges a SNSF mobility fellowship (177732). 
%

\appendix
\section{Numerical simulations}\label{section:theory} 
For the numerical simulations presented in the main text, we have taken 600 discrete bins for Larmor frequency distribution $\delta_j$ linearly spaced between $-5\kappa$ and $5\kappa$. Furthermore, spin linewidths are much larger than $\kappa$ and Rabi frequency, so we assume a constant spin distribution $\rho_\delta$. The distribution in coupling strength is incorporated by taking 150 bins of $g$ values, again linearly spaced between the maximum and the minimum value. The coupling strength distribution $\rho_g (g)$ is determined using a finite-element simulation of the magnetic field profile $\delta B_1(r)$ around the inductance (using COMSOL) and also the knowledge of the implantation profile below the inductor (measured by Secondary Ions Mass Spectroscopy). The minimum $g$ is determined by the lateral size of the box in the COMSOL simulation, which we choose to be four times the width of the inductor while centered around the inductor. These boxes are shown in the Fig.~\ref{fig:setup}(d-f). Since all measurements are done at low temperatures 20~mK and at large frequencies $\omega_0/2\pi \sim 7$~GHz, we take equilibrium polarization to be 1. However, because of finite repetition rate $\gamma_\text{rep}$, the initial conditions for the simulations are set by $S^{(j)}_z(t=0)=1-\text{exp}[-1/(T^{(j)}_1 \gamma_\text{rep})],~S^{(j)}_x(0) = S^{(j)}_y(0) =0$.

\end{document}